\documentclass[traditabstract, letter]{aa}
\usepackage{natbib}
\usepackage{txfonts}
\usepackage{graphicx}
\titlerunning{Convection and Cepheid period jitter}
\authorrunning{H.~R. Neilson \& R. Ignace}

\begin{document}

\title{Convection, granulation and period jitter in classical Cepheids}

\author{Hilding R. Neilson \and Richard Ignace \inst{1}}
\institute{
   Department of Physics \& Astronomy, East Tennessee State University, Box 70652, Johnson City, TN 37614 USA
   \email{neilsonh@etsu.edu}
  }

\date{}

\abstract{
Analyses of recent observations of the sole classical Cepheid in the Kepler field, V1154~Cygni, found random changes  of about 30~minutes in the pulsation period.  These period changes challenge standard theories of pulsation and evolution because the period change is non-secular, and explaining this period jitter is necessary for understanding stellar evolution and the role of Cepheids as precise standard candles.  We suggest that convection and convective hot spots can explain the observed period jitter.  Convective hot spots alter the timing of flux maximum and minimum in the Cepheid light curve, hence change the measured pulsation period. We present a model of random hot spots that generate a localized flux excess that perturbs the Cepheid light curve and consequently the pulsation period which is consistent with the observed jitter.  This result demonstrates how important understanding convection is for modeling Cepheid stellar structure and evolution, how convection determines the red edge of the instability strip, and just how sensitive Cepheid light curves are to atmospheric physics. 
}
\keywords{convection --- stars: late-type --- starspots --- stars: variables: Cepheids --- stars: individual: V1154 Cyg}
\maketitle

\section{Introduction}
Classical Cepheids are arguably archetypical standard candles since the discovery of the Cepheid period-luminosity relation or Leavitt law more than a century ago \citep{Leavitt1907}.  Because of the brightness of Cepheids, $\log L/L_\odot = 3$ -- $4.5$, and the fact that brightness is correlated with the pulsation period, Cepheids are powerful tools for measuring distances to galaxies \citep[e.g.][]{Gerke2011,Gieren2013} and even the Hubble constant \citep{Hubble1929, Freedman2001, Riess2011, Freedman2012}. An additional strength of Cepheids as standard candles is their consistent and precisely measured pulsation periods.

Cepheid pulsation periods are also observed to change over time. \cite{Eddington1919} presented the first measurements of secular period change in the prototype $\delta$ Cephei to confirm previous arguments that period change is a result of a central energy source that is not gravitational contraction.  It is now known that the rate of period change is a direct measure of stellar evolution \citep[e.g.][]{Struve1959} and can test the details of Cepheid evolution, especially Cepheid mass loss \citep{Neilson2012a, Neilson2012b}.  The measured changes of pulsation periods presented by \cite{Turner2006} are secular and slowly changing over many years and raise many question such as the nature of the evolution of the nearest Cepheid Polaris \citep{Neilson2012a, Turner2013a, vanLeeuwen2013}.

New observations are complicating this picture of secular evolution and secular period change. \cite{Derekas2012} presented Kepler observations of V1154~Cygni where they found that the pulsation period varies by about 0.015 - 0.02 day from its standard pulsation period of about 4.9 days, that is, $\Delta P/P \le 1\%$.  This period jitter is unexpected and presents challenges for detecting binary companions as well as for understanding the physics of Cepheids. \cite{Evans2013} also found similar period jitter from MOST observations of two other Cepheids and that there apparently is more jitter for first-overtone Cepheids. \cite{Derekas2012} made several suggestions to explain this jitter, including an instability in the pulsation period itself, or possibly convective granulation in Cepheids, analogous to convective cells observed in red supergiant stars like Betelgeuse \citep{Gray2008,Haubois2009}.  Convection is known to be important for pulsation in Cepheids, because it defines the red edge of the instability strip \citep[e.g.][]{Yecko1998, Bono1999}.

Convection is a particularly arduous topic to explore in classical Cepheids and other stars as it is intrinsically a multidimensional problem. Typically, Cepheid structures are computed in one-dimension with various treatments of convection and numerous free parameters \citep[e.g.][]{Buchler2002, Smolec2008}. New efforts are considering multidimensional simulations of Cepheid structures. \cite{Gastine2011} computed two-dimensional stellar models to explore the interaction between convection and pulsation in a Cepheid.   However, their model also assumes free parameters akin to one-dimensional models. Similarly, \cite{Geroux2013} presented two-dimensional radiation hydrodynamic models of convection in RR Lyrae stars and found that simulated light curves matched observational light curves. \cite{Mundprecht2012} computed three-dimensional Cepheid models that hint at the presence of granulation.  While these models are promising, they are also computationally expensive, requiring both high spacial and temporal resolution, even more so than that required for simulating convection in red supergiant stars \citep{Freytag2013}.

In this work, we construct a model for convective hot spots that vary as a function of time, number, and size, producing a radiant contribution that is superimposed on a Cepheid light curve.  In Sect.~2, we describe the model for convective spots and pulsation.  In Sect.~3, we present results for a fundamental-mode model light curve, and results for a first-overtone model light curve in Sect.~4.  We summarize our results and discuss implications in Sect.~5.
 
\section{Modeling star spots}
We hypothesize that convection can cause period jitter like that observed for V1154 Cyg.  The observed period jitter is a random-like shift in the timing of the brightness minimum and maximum. Convective hot spots lead to granulation-like noise that can vary the precise timing of the brightness minimum and maximum.  Similarly, convective hot spots have been observed for the red supergiant Betelgeuse \citep{Haubois2009}. Granulation signatures have also been observed in radial velocity data for Kepler target stars, and it has been shown that the radial velocity jitter correlates with stellar fundamental parameters \citep{Bastien2014}.

\begin{figure}[t]
\begin{center}
\includegraphics[width=0.47\textwidth]{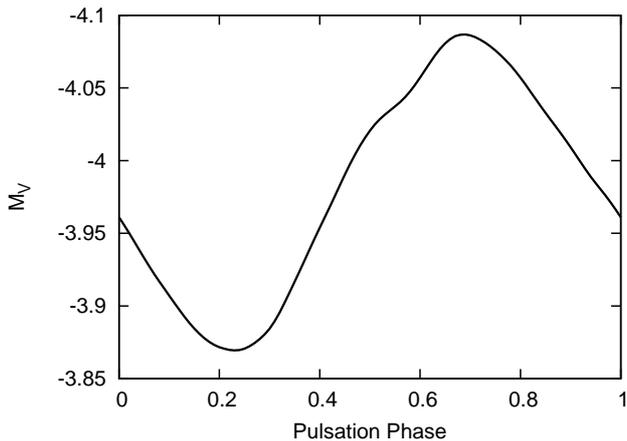}
\end{center}
\caption{$V$-band light curve in absolute magnitudes for a 10-day period Cepheid where the light curve is described by the \cite{Pejcha2012} prescription.}
\label{fund_light}
\end{figure}

%Consider a fundamental-mode Cepheid with a 10~day pulsation period whose light, radius and temperature variation is prescribed by \cite{Pejcha2012}, with a light curve is shown in Fig.~\ref{fund_light}. 
In Fig.~\ref{fund_light} we show the light curve prescribed by \cite{Pejcha2012} for a fundamental-mode Cepheid with a 10 d pulsation period. These authors derived the variations of all the relevant physical quantities along a pulsation cycle 
through a global fit to an extended database of observed pulsation properties for Galactic and Magellanic Cepheids \cite[see][for details]{Pejcha2012}.
If that Cepheid has a period jitter equivalent to that of V1154 Cyg, then the period would vary by about 1.2~hours.  This corresponds to a phase difference of $\Delta \phi = 0.005$, and a change of magnitude, $\Delta M_V < 0.001$.  Therefore, if convective flux variation is about the same order of magnitude, then convective hot spots can explain the period jitter. Furthermore, if the Cepheid is an s-Cepheid or first-overtone Cepheid with a low-amplitude sinusoidal light curve, then similar convective hot spots will cause even stronger period variation because of the more shallow slope of the light curve near flux minimum and maximum. \cite{Evans2013} noted that period variations appear to be consistent for a number of cycles. Therefore, this hypothesis requires that the convective spot lifetime must be about that long as well, especially since a Cepheid's rotation rate is $\le 10~$km~s$^{-1}$ \citep{Bersier1996}.  

% How convection can lead to a period jitter
% Period jitter is a time shift in the location of light maximum and minimum
% Consider a standard light curve with P = 10 d and a period jitter of 1 hour, scaled up from the observations
% Over that time span the flux change is Delta M/dphi = xxx
% Therefore if convective flux variation is  of the same order then convection can explain the observed jitter

We constructed a toy model to explore the role of convective hot spots in perturbing the light-curve structures of classical Cepheid stars. This toy model starts by assuming a Cepheid light curve, either a sinusoidal function for first-overtone pulsation or the model provided by \cite{Pejcha2012} for fundamental-mode pulsation. The second step is to then include light fluctuations from convective spots into these light curves at a given phase.

We treated convection and convective spot formation as a random process where at any given time spots randomly form, have a pre-determined lifetime, and fade. There are a random number of hot spots at any given time, based on the assumption there are so many spots over the total number of pulsation cycles considered, with a given temperature that is higher than the stellar effective temperature.  It must be noted that spot size, spot number, and spot temperature are degenerate parameters when computing their contributions to the total stellar flux, but our goal here is to devise a simple test to understand whether convective spots can explain the observed period jitter in Cepheids.

To compute models for this work, we considered three parameters for convective spots: the opening angle of a spot, the relative spot temperature, and the probability of a spot forming.  We assumed the spot lifetime is about one-tenth of a pulsation period, but it can be much longer without significantly changing the total amount of period jitter, only the correlation of the period shift from one cycle to the next.  For instance, if the cell lifetime is much longer than the pulsation period, then the period will appear to decrease for a number of cycles before returning to the original value and likewise increase.   For this work, we considered the spot opening-angle to vary from about $4$ to $7^{\circ}$, the temperature to be about 3 to 6\% of the stellar effective temperature, and 20\% probability of there being at least one star spot at any time, where we ignored the role of rotation. The opening angle and temperature variation are arbitrary choices, limited only by the measured convective cell properties of the red supergiant Betelgeuse \citep{Haubois2009}.  We consider these values to be intermediate values for stars that are hotter and more compact than Betelgeuse.  

\section{Results for fundamental-mode pulsation}

\begin{figure*}[t]
\begin{center}
\includegraphics[width=0.47\textwidth]{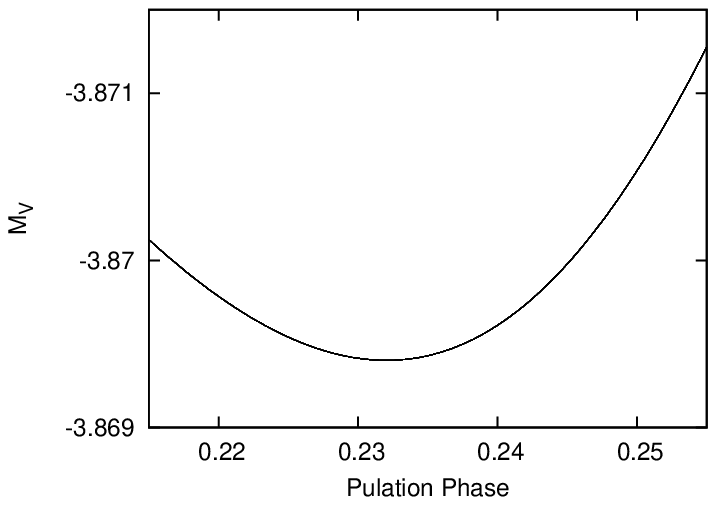}\includegraphics[width=0.47\textwidth]{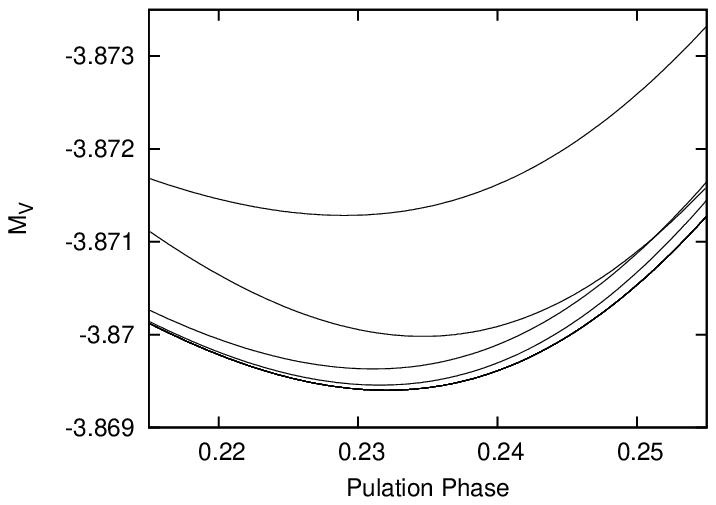}
\end{center}
\caption{Variation of a Cepheid light curve due to convective hot spots for two cases near flux minimum, where the light curves are folded and the pulsation phase is defined by the initial unperturbed light curve.  (Left) The case where spots have an opening angle of $4^\circ$ and $T_{\rm{spot}} = 1.03T_{\rm{eff}}$ and (right) where spots  have an opening angle  of $7^\circ$ and $T_{\rm{spot}} = 1.06T_{\rm{eff}}$.}
\label{fund_small}
\end{figure*}

We computed a series of model light curves for the fundamental-mode Cepheid with a 10-day pulsation period whose light, radius, and temperature variation is prescribed by \cite{Pejcha2012}.   The corresponding light curve is shown in Fig.~\ref{fund_light}.  We computed ten pulsation cycles where random hot spots were superimposed.  The first case is for smaller spots with an opening angle of $4^\circ$ and a relative temperature of $1.03 \times T_{\rm{eff}}$.  We computed a second case for a larger angle of $7^\circ$ and a temperature of $1.06\times T_{\rm{eff}}$.  Both cases are shown in Fig.~\ref{fund_small}.

The computation of the first case suggests that convective spots cannot be too small or too cool.  If the effective temperature at flux minimum is about $5280~$K, then the spot temperature is about 5440~K.  Moreover, the spot size is small, an opening angle of 4$^\circ$ suggests a relative surface area of $< 0.5\%$ of the stellar disk, hence this only enhances the luminosity by about $0.003\%$ at the flux minimum, much lower than the scatter of Kepler observations.  Many spots would have to be created at the same time to drive period jitter in this case.  

The results for the second case, however, are consistent with observed period jitter.  The phase where minimum flux occurs varies by $\Delta \phi \approx 0.01$ for a 10-day-period Cepheid. This translates into a period jitter of about 2.4 hours, about twice as long as the period jitter extrapolated from V1154 Cyg. Therefore, the observed period jitter can be explained by convective cells, but are these convective spot properties possible?

\cite{Haubois2009} presented interferometric measurements of the red supergiant Betelgeuse and found evidence for large convective spots with an opening angle that is about 1/4 of the stellar diameter and a temperature that is about $1.14\times T_{\rm{eff}}$.  These values are much higher than those necessary to explain Cepheid period jitter.  Numerical simulations of red supergiants also have similar convective cell properties as those observed \citep{Chivassa2010}. 
Hence, we have shown that convective granulation in Cepheids explains the period jitter detected by \cite{Derekas2012}.

\section{Results for first-overtone pulsation}
\begin{figure}[t]
\begin{center}
\includegraphics[width=0.47\textwidth]{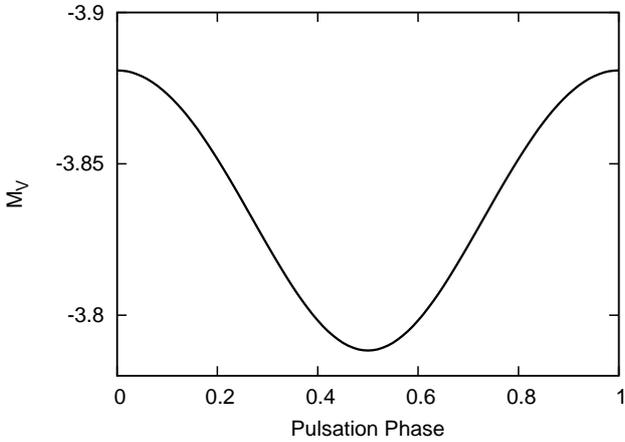}
\end{center}
\caption{$V$-band light curve for a small amplitude Cepheid where the light curve is sinusoidal.}
\label{over_fig}
\end{figure}

We repeated our calculations for first-overtone Cepheids or s-Cepheids.  These stars typically have more sinusoidal-like light curves, for instance, Polaris and FF Aql \citep{Neilson2012a, Turner2013b,Turner2013a}.  Because of the shape of the light curve, determining the time of light minimum is more difficult than for a fundamental-mode Cepheid.  A typical sinusoidal light curve is shown in Fig.~\ref{over_fig} with a mean bolometric luminosity, $\log L/L_\odot = 3$ and $T_{\rm{eff}} = 6200$~K, similar to FF Aql, but less luminous.

\begin{figure*}
\begin{center}
\includegraphics[width=0.47\textwidth]{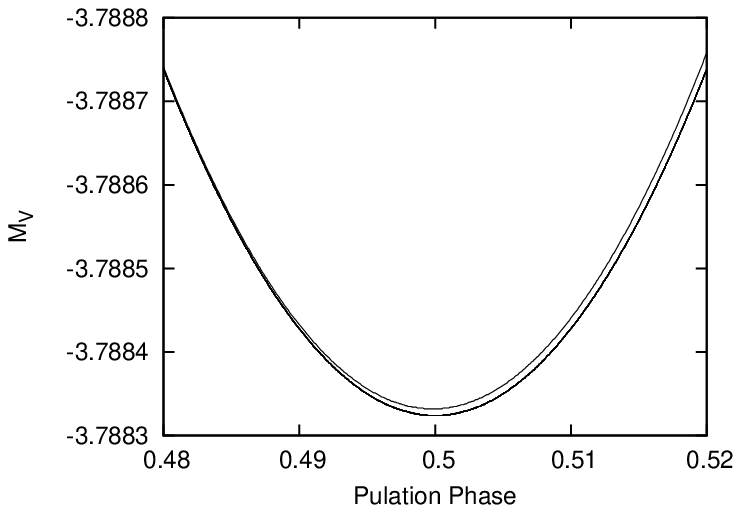}\includegraphics[width=0.47\textwidth]{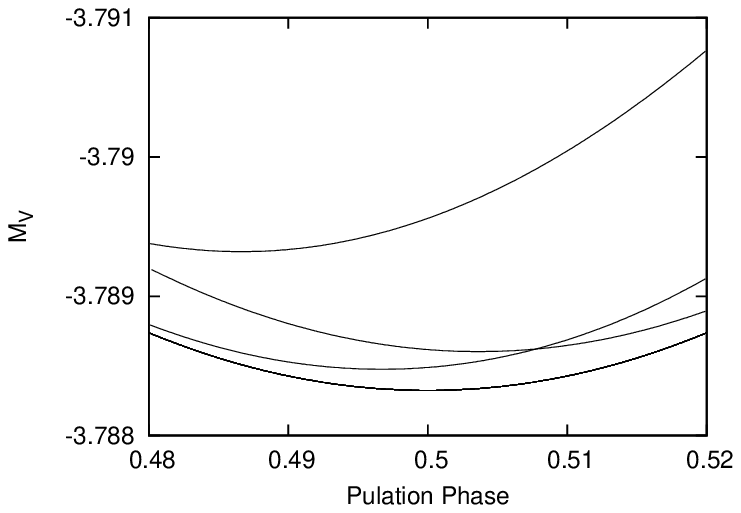}
\end{center}
\caption{Variation of a first-overtone Cepheid (or s-Cepheid) light curve due to convective hot spots for two cases near flux minimum.  (Left) The case where spots have an opening angle of $4^\circ$ and $T_{\rm{spot}} = 1.03T_{\rm{eff}}$ and (right) where spots  have an opening angle of $7^\circ$ and $T_{\rm{spot}} = 1.06T_{\rm{eff}}$.}
\label{first_small}
\end{figure*}

The same spot parameters as those chosen in the previous section were assumed here, and the resulting light-curve variations are shown in Fig.~\ref{first_small}.  The results for a first-overtone Cepheid are similar to the results for the fundamental-mode Cepheid for the case where star spots have a small opening angle and low temperature excesses.  However, the case for which the starspots are larger and hotter, the resulting period jitter and flux variation differs.   The flux variation for the first-overtone Cepheid is about $\Delta M_V \approx 0.001$ magnitudes, whereas for the fundamental-mode Cepheid it is $\Delta M_V \approx 0.002$ magnitudes.  There is more period jitter for the first-overtone Cepheid with $\Delta \phi \approx 0.02$, about twice as much as for the fundamental-mode Cepheid.  However, it should be noted that our model for the first-overtone Cepheid assumed an arbitrary pulsation period.

The amount of period jitter found in our first-overtone Cepheid model is equivalent to a time variation of about 2~hours if we assume the pulsation period is that of FF Aql (P = 4.47~days). This period is much shorter than the period of our fundamental-mode Cepheid, but does suggest that low-amplitude sinusoidal pulsation will show more relative period jitter. 

\section{Discussion}
The purpose of this work was to understand whether hot convective cells and convective variations might cause the previously observed period jitter \citep{Derekas2012}.  We computed random hot-spot variations given a number of free parameters such as spot size, spot temperature, spot lifetime, and the number of spots. We found flux variations ranging from $< 0.0001$ mag to about $0.002$ mag and shifts in light minimum up to $\Delta \phi = 0.01$ for a fundamental-mode Cepheid and $0.02$ for a first-overtone Cepheid.

These results are more extreme than that observed. The Kepler space telescope is able to resolve flux variations of $0.0001$ mag and a period jitter of about $0.005$, lower than the limits we computed.  However, our model is simple and assumes many degenerate free parameters. The model also adopts a Cepheid light curve and then superimposes convective spot fluxes.  In reality, the Cepheid light curve that was constructed from observations \citep{Pejcha2012} would contain the average convective flux, so variation due to spots would both increase and decrease the relative flux, hence we are arguably overestimating the flux variation.  Similarly, small changes in our free parameters, such as increasing the number of spots, including limb-darkening and even rotation (albeit slow) could also play a role.

The hypothesis that period jitter is caused by convective hot spots implies that period jitter is itself a function of pulsation period.  We computed the period jitter for a 10-day-period Cepheid, but the period was assumed only for the purpose of constructing the initial light curve. In principle, the amount of relative period jitter depends primarily on the size and temperature of the hot spots. Convection is a function of effective temperature and becomes stronger at cooler effective temperatures.  This suggests that hotter Cepheids would have smaller hot spots and a weaker relative temperature increase (i.e., $T_{\rm{spot}}/T_{\rm{eff}}$) than cooler Cepheids.  Because there is a linear relation between the effective temperature and pulsation period, as suggested by the Cepheid period-color relation \citep[e.g.][]{Tammann2003}, there will be more relative jitter at longer periods.  This suggests that long-period Cepheids, such as $l$~Car and RS~Pup, will have more relative period jitter than short-period Cepheids such as $\delta$~Cep and  V1154~Cyg. Furthermore, this period jitter may complicate distance estimates from ultra-long-period Cepheids ($P>80$~days) \citep{Bird2009, Fiorentino2012}.

On the other hand, it may be possible to use the observed period jitter to constrain fundamental stellar parameters similar to the method suggested by \cite{Bastien2014}.  As convection and convective instability is a function of effective temperature and gravity, then so must be the relative amount of jitter.  One can then combine the amount of period jitter with the average period to help calibrate the Cepheid period-luminosity relation, and fundamental properties for constraining stellar evolution models and multidimensional simulations.

Convective hot spots in Cepheids can be observed by various methods such as interferometric and polarization measurements.  More robust interferometric measurements of the third and fourth lobes of the visibility curve of a Cepheid would constrain perturbations in the intensity profile due to star spots, analogous to the observations by \cite{Haubois2009}.  Similarly, linear polarization measurements will constrain asymmetric structures on the stellar surface \citep{Schwarz1984, Clarke1984}. The combination of the two observations will help constrain the presence and evolution of convective cells and confirm the toy model presented in this work.

While our model is simple, the results are robust enough to conclude that convection and convective cells explain the observed period jitter.  However, detailed tests require three-dimensional radiation hydrodynamic simulations of Cepheid atmospheres and the interaction between pulsation and convection.  \cite{Fokin1996} computed one-dimensional RHD models for $\delta$ Cep and found that numerous shocks propagated throughout the photosphere, reaching velocities of up to three times the sound speed, hence adding another challenge to computing multidimensional models. Even with these challenges, new three-dimensional models of Cepheids are being computed, and this new field is still in its infancy \citep{Mundprecht2012}.  In the future, it will be possible to compute grids of pulsating Cepheid atmospheres and potentially verify this work.

% Kepler < 10^-4 mag and delta phi ~0.005
% we find up delta phi ~0.01 and ~10-3 mags
%But our model is toy
%assumes the light curve is the minimum amount of light when it should be a "mean", representing the averaging convective cell flux. 
% This would lead to smaller flux variations but can maintain similar period jitter
% Similar having more spots at one time and including the role of limb darkening and rotation will also affect the resultant jitter 
%Our model is simple, but to test the hypothesis in more detail requires 3-D rad hydro atmospheres
% these computations are in their infancy, and requires more intensive computing that the grids of non-pulsating atmospheres as pulsation requires finer time resolution especially with the presence of shocks
\acknowledgements

This work has been supported by a research grant from the NSF (AST-0807664). The authors would also like to thank Nancy Evans for conversations that have helped this work.

\bibliographystyle{aa} % style aa.bst

\bibliography{jitter} % reference ld_p2.bib

\end{document}